\documentstyle[twoside,fleqn,qcdstyle]{article}

\begin{document}

\title{On General Axial Gauges for QCD
\thanks{Talk presented by DFL at QCD'98, Montpellier, July 1998. To be published in the proceedings.}
}

\author{
Daniel F. Litim${}^{a}$ and Jan M. Pawlowski${}^{b}$\\[2ex]
${}^a$Departament ECM \& IFAE,
Facultat de F\'{\i}sica, Universitat de Barcelona\\
$\, \;$Diagonal 647, E-08028 Barcelona, Spain.\\[2ex]
${}^b$Dublin Institute for Advanced Studies,
10 Burlington Road, Dublin 4, Ireland.\\[2ex]
}

\begin{abstract}
General Axial Gauges within a perturbative approach to QCD are plagued by 'spurious' propagator singularities. Their regularisation has to face major conceptual and technical problems. We show that this obstacle is naturally absent within a Wilsonian or 'Exact' Renormalisation Group approach and explain why this is so. The axial gauge turns out to be a fixed point under the flow, and the universal 1-loop running of the gauge coupling is computed.
\vspace*{-8.2cm}
\begin{flushright}
{\normalsize ECM-UB-PF-98-21 \\  DIAS-STP-98-09}
\end{flushright}
\vspace*{6.7cm}
\end{abstract}
\maketitle

\section{Introduction}
General axial gauges (GAGs) \cite{kummer} have been studied extensively within the few last decades \cite{axialbook}. Their main feature is  the decoupling of the ghost sector since the Fadeev-Popov determinant no longer depends on the gauge field. This reduces the number of possible vertices significantly, and is a major simplification from a technical point of view. Furthermore, unlike covariant gauges, GAGs do not suffer from the possible presence of Gribov copies. However, they need the introduction of an additional Lorentz vector  $n_\mu$. As a consequence, one has to cope with more structure functions per vertex due to the richer  Lorentz structure, which partly spoils the above advantages. 
The key obstacles, however, are the so-called 'spurious' divergences of the perturbative propagator. These singularities are of the form $(np)^{-\alpha}, \alpha\ge 1$ and have to be regulated separately. They can be interpreted as a remnant of the gauge fixing being incomplete, which leaves the quadratic part of the gauge-fixed action 'not fully' invertible. 
In the sequel, we will argue that the Wilsonian (or 'Exact') Renormalisation Group (RG) \cite{Wilson}-\cite{EAA} is an appropriate alternative framework for this problem. It is self-contained, it allows for non-perturbative approximations, and -most importantly- the 'spurious' divergences are naturally absent.

\section{The 'spurious' propagator singularities}
To begin with, let us introduce the gauge fixing term for a GAG as 
\beq \label{gf}
S_{\rm gf}=\01{2}\int d^dx \ n_\mu A^a_\mu\ \01{\xi n^2}\ n_\nu A^a_\nu 
\eeq 
in a  $d$-dimensional Euclidean space. It is quadratic in the gauge field, and the gauge fixing parameter $\xi$ with mass dimension $-2$ may even be momentum dependent. For the following considerations, fermions act as spectators. Therefore we can restrict ourselves to the purely gluonic part of the action, which is 
\bea S_A&=&\01{4}\int d^dx\ F_{\mu\nu}^a F_{\mu\nu}^a \\ 
F_{\mu\nu}^a&=&\partial_\mu A^a_\nu-\partial_\nu A^a_\mu + g f^a_{\ \   bc} A^b_\mu A^c_\_\nu \nonumber\\
D^{ab}_\mu&=&\delta^{ab}\partial_\mu + g f^{acb}A^c_\mu, \ \ [t^b,t^c]={f_a}^{bc}t^a. \nonumber
\eea 
The perturbative gluonic propagator $P_{\mu\nu}$ can be decomposed on a basis of symmetric Lorentz tensors as
\beq
a_1\frac{\delta_{\mu\nu}}{p^2}+a_2 \frac{p_\mu 
p_\nu}{p^4} +a_3
\frac{n_\mu p_\nu +n_\nu p_\mu}{p^2 (np)}+a_4 \frac{n_\mu
  n_\nu}{n^2 p^2}.
\label{prop_k}
\eeq 
The dimensionless coefficients $a_{1,...,4}$  as obtained from 
the gauge-fixed classical action $S=S_A+S_{\rm gf}$ 
are
\bea 
a_1&=&1\label{a1}\\ a_2&=&(1+\xi p^2)/s^2\\
a_3&=&-1\\ a_4&=&0\label{a4}\\ s^2&=&(np)^2/n^2p^2.  
\eea 
$P_{\mu\nu}$ displays the usual IR poles proportional to $1/p^2$. In addition, we observe divergences for momenta orthogonal to $n_\mu$. These poles appear explicitly up to second order in $1/np$ and can even be of higher order for certain $np$-dependent choices of $\xi$. For the planar gauge, $\xi p^2 =-1$, the spurious divergences appear only up to first order. 
Within perturbation theory, quantum corrections are computed in a loop expansion, with the bare propagator $P_{\mu\nu}$ appearing in the loop integrals. The question about how to regularise $P_{\mu\nu}$ such as to allow for a consistent loop expansion stimulated extensive investigations (see \cite{axialbook} and references therein). The intricacies concerning these regularisations partly spoil the advantage of having fewer diagrams to calculate. Furthermore, it is not fully established whether the proposed regularisation schemes are consistent at higher loop order.
\section{The 'Exact' Renormalisation Group}
A very promising tool for studying both perturbative and non-perturbative aspects of quantum field theories is given by the Wilsonian or 'Exact' Renormalisation Group \cite{Wilson}-\cite{EAA}. Its formulation, as presented for example in \cite{EAA}, relies on the presence of a scale dependent regulator term $\Delta_k S$. It is aimed at (i) regulating possible IR divergences and (ii) integrating-out momentum shells of quantum fluctuations around $p^2\approx k^2$. The scale parameter $k$ can be interpreted as a coarse-graining scale. We use
\beq \Delta_k S[A] = \frac{1}{2} \int \frac{d^dp}{(2\pi)^d}\ A^a_\mu
R^{ab}_{k,\mu\nu}(p) A^b_\nu,
\label{cut-off1}
\eeq 
which is quadratic in the fields. The related scale dependent effective action $\Gamma_k[A]$ can be shown to interpolate between the gauge-fixed classical action as given at some initial UV scale $\Lambda$,  $\Gamma_{k\to\Lambda}=S_A[A]+S_{\rm gf}[A]$,  and the full quantum effective action $\Gamma_{k\to 0}\equiv \Gamma$. It obeys the flow equation 
\beq\label{flow} 
\partial_t\Gamma_k[A]=\frac{1}{2}{\rm Tr}
\left\{G_k[A] \frac{\partial R_k}{\partial t}\right\}. 
\eeq 
Here, the trace  sums over all momenta and indices, $t=\ln k$, and
\beqa\label{prop}
G_{k,\mu\nu}^{ab}=\di \left(\frac{\delta^2\Gamma_k[A]}{\delta
  A_\mu^a\delta A_\nu^b}+R_{k,\mu\nu}^{ab} \right)^{-1}
\eeqa 
denotes the full ({\it i.e.}~field-dependent, regularised) propagator. A typical regulator with the required properties is given by
\bea
R^{ab}_{k,\mu\nu}(p)&=&\delta^{ab}\delta_{\mu\nu}p^2 \ r(p^2/k^2)
\label{cut-off2}\\
r(p^2/k^2)&=&({\exp (p^2/k^2) - 1})^{-1}.
\eea
Note that for a given scale $k$ only momentum fluctuations with momenta about $k$ can contribute in \eq{flow}. This is precisely due to the insertion of the operator $\partial R_k/\partial t$, which is peaked about $p^2\approx k^2$ and (exponentially) suppressed elsewhere. It acts like a 'mode window'. Integrating \eq{flow} w.r.t.~$k$ corresponds to summing-up quantum fluctuations. The problem of computing the IR physics starting with the short distance behaviour of the theory -{\it i.e.}~with some UV initial action $\Gamma_\Lambda$- reduces to the problem of integrating the flow equation \eq{flow} \cite{gag},\cite{we}.
 
What have we gained as opposed to the perturbative approach?
Let us consider the propagator $P_{k,\mu\nu}$ obtained from $S_k=S_A+S_{\rm gf}+\Delta_k S$. Decomposing $P_{k,\mu\nu}$ as in \eq{prop_k}, we obtain a set of dimensionless coefficients that depend on the regulator function, momenta, $n_\mu$, and the gauge fixing parameter in the following ways:
\bea 
\di a_1 &=&1/(1+r)\label{ak1}\\
\di a_2 &=&[1 + \xi p^2 (1+r)]/z \\
\di a_3 &=&- s^2/z\\
\di a_4 &=& -r/z\label{ak4}\\
z&=&(1+r)[ s^2 +(1+\xi p^2 (1+r))r].  \eea
By construction, $P_{k,\mu\nu}$ has the following limits
\bea \di
\lim_{k\rightarrow 0}P_{k,\mu\nu} &=& P_{\mu\nu}\label{lim1}\\ 
\di 
\lim_{p^2\rightarrow 0}P_{k,\mu\nu} &=&
  \frac{\delta_{\mu\nu}}{k^2}+\frac{1}{1+\xi k^2}\frac{n_\mu n_\nu}{k^2n^2}
 \label{lim2} \\
\di \lim_{k\rightarrow \infty}P_{k,\mu\nu}&=&  0 \label{lim3}
\eea 
They reflect, respectively, the vanishing of the regulator term for $k=0$, the IR finiteness for $k>0$, and the non-propagation of fluctuations for large scales. The important observation is now that even the limit of $P_{k,\mu\nu}$ for $(np)\to 0, k\neq 0$ remains finite, unlike its divergent perturbative counterpart $P_{\mu\nu}$! This holds true  for an arbitrary choice of $\xi$ and leads to 
\bea
P_{k,\mu\nu} & =& \di \frac{1}{1+r}\frac{\delta_{\mu\nu}}{p^2}+
\frac{1}{(1+r)r}\frac{p_\mu p_\nu}{p^4}\nonumber \\
&&-\di\frac{1}{(1+r)(1+p^2\xi (1+r))} \frac{n_\mu n_\nu}{n^2
  p^2}\label{prop_np}.  
\eea
Thus $P_{k,\mu\nu}$  is perfectly well-behaved and finite for all momenta $p$. It is noteworthy that the 'spurious' divergences are absent as soon as the infra-red behaviour of the propagator is under control.

\section{Gauge invariance}
The delicate question to raise concerns gauge invariance of physical Green functions \cite{gag},\cite{we}. Merely adding the quadratic term (\ref{cut-off1}) to the action seems to be in conflict with gauge invariance. 
In the present context, the requirement of gauge invariance translates into a {\it modified} Ward Identity (mWI) 
\bea \di 
  0&=& \di D_\mu^{ab}(x)\frac{\delta
 \Gamma_k[A]}{\delta A^b_\mu(x)} \di-n_\mu D_\mu^{ab}(x)\ 
\frac{1}{n^2\xi}
n_\nu A_\nu^b(x) \nonumber \\[2ex]&-&\di g \int \! d^dy\ f^{abc}\left(\frac{n_\mu
  n_\nu\delta^{cd}}{n^2\xi} +R^{cd}_{k,\mu\nu}\right)
  G_{k,\nu\mu}^{db},
\label{mWIk}\eea
where the 'modification' stems from the regulator  term $R_k$. The mWI \eq{mWIk} is compatible with the flow \eq{flow} \cite{gag}. This implies that any functional $\Gamma_{\Lambda}$ which is a solution of the mWI at some scale $\Lambda$ will remain a solution for any $k<\Lambda$, if $\Gamma_k$ evolves according to the flow equation. For $k\to 0$, \eq{mWIk} reduces to the usual WI. This establishes  that the {\it physical} Green functions (obtained from $\Gamma_k$ after integrating-out all momentum fluctuations down to $k=0$) do obey the usual WI, {\it i.e.}~gauge invariance is maintained. 

\section{Applications}
We shall use the above to show both that the axial gauge is a fixed point under the flow, and that the universal 1-loop $\beta$-function is recovered. 

It was shown non-perturbatively that the axial gauge $(\xi =0)$ is uniquely defined \cite{gag}. This serves as the starting point to prove that the axial gauge is a fixed point under the Wilsonian flow. Our proof is based on the modified Ward Identity (mWI), and the finiteness of the flow equation. No assumptions on the functional form of the full quantum effective action have to be made. Without loss of generality, we can choose $\xi$ as momentum independent, in which case the mWI \eq{mWIk} simplifies to
\bea  
\di 0&=&\di D_\mu^{ab}(x)\frac{\delta
  \Gamma_k[A]}{\delta A^b_\mu(x)} \di-\frac{1}{n^2\xi} 
n_\mu\partial^x_\mu\ n_\nu A^a_\nu (x)\di \nonumber \\ &&
-g \int d^dy\ f^{abc}R^{cd}_{k,\mu\nu}
(x,y) G_{k,\nu\mu}^{db}(y,x).  
\label{mWI}\eea 
Note that $\xi$ enters the mWI both explicitly and implicitly. The $\xi$ appearing explicitly corresponds to the choice of $\xi$ at some initial scale $\Lambda$, $\xi\equiv\xi(\Lambda)$. An implicit dependence occurs through the dependence of $\Gamma_k$ on the scale dependent $\xi(k)$. Let us choose $\xi(\Lambda)=0$ with $\Gamma_\Lambda$ solving (\ref{mWI}) and {\em assume} that $\xi(k)\neq 0$ for some $k<\Lambda$. This means in particular that $\Gamma_k$ will no longer contain a singular term $\sim 1/\xi$. Thus the only singular term appearing in the mWI is the term explicitly proportional to $1/\xi$. One can always find an $A^a$ such that $n_\mu\partial_\mu\ n_\nu A_\nu^a$ does not vanish. Therefore $\Gamma_k$ with $\xi(k)\neq 0$ cannot be a solution of (\ref{mWI}) for $\xi(\Lambda)=0$. But this is in contradiction with the compatibility of the flow equation and the mWI, which  implies that $\Gamma_k$ solves the mWI. It follows that $\xi(k)=0$ for $\xi(\Lambda)=0$. Hence the axial gauge 
is indeed a fixed point of the flow equation.

Finally, we summarise the key steps for the computation of the universal  1-loop $\beta$-function within the Wilsonian RG approach \cite{1-loop}. This is an interesting check for two reasons. Firstly, the flow equation \eq{flow} does depend explicitly on the regulator function, while the corresponding 1-loop $\beta$-function -known to be universal- has to be independent of it. Secondly, it is known that  the $\beta$-function -within a perturbative computation-  receives contributions from the (regularised) 'spurious' divergences already at the 1-loop level. In the present approach, the theory is finite by construction, and in particular there are no such contributions from 'spurious' divergences. Thus, it is important to see how the universal $\beta$-function is obtained in the present approach. 

To that end, we set $d=4$ and fix the gauge to the axial one, $\xi= 0$. The running gauge coupling  is introduced as $g_k=g Z_{g,k}$. At the 1-loop level, the scale dependent wave function renormalisation of the gauge coupling $Z_{g,k}$ is related to that of the field strength, $Z_{F,k}$, through 
\beq
Z^2_{g,k}=Z^{-1}_{F,k}.
\eeq 
This follows from the mWI \cite{we}. We use the Ansatz 
\beq \label{ansatz}
\Gamma_k= \014\int d^4x\ Z_{F,k}\, F_{\mu\nu}^a F_{\mu\nu}^a 
+ S_{\rm gf}[Z_{F,k}^{1/2}A]
\eeq
Inserting \eq{ansatz} into \eq{flow}, we obtain the 1-loop flow for $Z_{F,k}$ by projecting on the $F^2$-term in the flow equation. This procedure is self-consistent at the 1-loop level. The traces involved in \eq{flow} can be calculated using heat kernel methods \cite{1-loop}, with the result
\beq\label{runZ}
\partial_t \ln Z_{F,k} = \0{11N}{24\pi^2} g^2 + {\cal O}(g^4).
\eeq 
 The $\beta$-function follows immediately from \eq{runZ} as  
\beq
\beta_{g^2}\equiv \partial_t g^2_k = -\0{11N}{24\pi^2}g_k^4 + {\cal O}(g_k^6),
\eeq
{\it i.e.}~the well-known universal 1-loop result.
\section{Conclusions}
We have presented a formulation of QCD based on an 'Exact' RG. This approach is self-consistent, and possible IR divergences are under control. Most importantly, gauge invariance of physical Green functions is maintained. The 'spurious' divergences of perturbation theory appear to be naturally absent, which is ultimately a consequence of the IR regulation of the theory. As a result, the axial gauge turns out to be a fixed point, and the universal 1-loop running of the gauge coupling is recovered.

This formulation is a good starting point for investigations beyond perturbation theory in particular. Typical candidates are theories where the Lorentz symmetry is naturally broken, {\it e.g.}~QCD at finite temperature, where the rest frame of the heat bath singles out a fixed Lorentz vector. Another interesting application concerns a computation of expectation values of Wilson loops, which serve as order parameters for confinement. We hope to report on these matters in the future.

\end{document}